\documentclass[aps, prd, twocolumn, notitlepage, nofootinbib]{revtex4-1}
\usepackage{amsfonts,amssymb,amsmath}
\usepackage{color}
\usepackage{graphicx}
\usepackage[normalem]{ulem}

\newcommand{\tcb}{\textcolor{blue}}

\newcommand{\ti}{\textit}
\newcommand{\nn}{\nonumber}

\newcommand{\be}{\begin{equation}}
\newcommand{\ee}{\end{equation}}
\newcommand{\ba}{\begin{eqnarray}}
\newcommand{\ea}{\end{eqnarray}}

\newcommand{\bi}{\begin{itemize}}
\newcommand{\ei}{\end{itemize}}
\newcommand{\bfig}{\begin{figure}}
\newcommand{\efig}{\end{figure}}
\newcommand{\bc}{\begin{center}}
\newcommand{\ec}{\end{center}}

\newcommand{\lb}{\left(}
\newcommand{\rb}{\right)}

\newcommand{\vv}{\vec{v}}

\newcommand{\vx}{\vec{x}}

\newcommand{\ra}{\rightarrow}

\begin{document}

\title{Dilatonic Imprints on Exact Gravitational Wave Signatures}

\author{Fiona McCarthy}
\email{fmccarthy@perimeterinstitute.ca}
\affiliation{Perimeter Institute, 31 Caroline St. N., Waterloo,
Ontario, N2L 2Y5, Canada}
\affiliation{Department of Physics and Astronomy, University of Waterloo, Waterloo, Ontario, Canada, N2L 3G1}

\author{David Kubiz\v n\'ak}
\email{dkubiznak@perimeterinstitute.ca}
\affiliation{Perimeter Institute, 31 Caroline St. N., Waterloo,
Ontario, N2L 2Y5, Canada}
\affiliation{Department of Physics and Astronomy, University of Waterloo, Waterloo, Ontario, Canada, N2L 3G1}

\author{Robert B. Mann}
\email{rbmann@uwaterloo.ca}
\affiliation{Perimeter Institute, 31 Caroline St. N., Waterloo,
Ontario, N2L 2Y5, Canada}
\affiliation{Department of Physics and Astronomy, University of Waterloo, Waterloo, Ontario, Canada, N2L 3G1}

\date{March 5, 2018}


\begin{abstract}
 By employing the moduli space approximation,
we analytically calculate the
gravitational wave signatures emitted upon the merger of two extremally charged dilatonic black holes.
We probe several values of the dilatonic coupling constant $a$, and find significant departures from the Einstein--Maxwell ($a=0$) counterpart studied in \cite{Camps:2017gxz}. For  (low energy) string theory black holes $(a=1)$ there are no coalescence orbits and only a memory effect is observed, whereas for an intermediate value of the coupling
$(a=1/\sqrt{3})$ the late-time merger signature  becomes exponentially suppressed, compared to the polynomial  decay in the $a=0$ case without a dilaton.  Such an imprint shows a clear difference between the case with and without a scalar field (as for example predicted by string theory) in black hole mergers.
\end{abstract}

\maketitle

\section{Introduction}

The great discovery made by LIGO on September 14, 2015 \cite{Abbott:2016blz} provided the first direct confirmation that strong gravitational
waves are  emitted in the process of the coalescence of two black holes.
The first event was for black holes of around 30 solar masses;
other discoveries soon followed and  gravitational waves have been now detected from several binary black hole mergers over a range of masses \cite{Abbott:2016nmj,Abbott:2017vtc,Abbott:2017oio,Abbott:2017gyy}. The most recently announced event is  from a neutron star-neutron star collision \cite{TheLIGOScientific:2017qsa},  marking the beginning of   multi messenger astronomy.

 To understand a black hole merger (or scattering) and the associated emission of gravitational waves is a complicated problem in which   strong field dynamical effects play an important role. For this reason, there is little hope for attacking this problem exactly, and various approximations \cite{Blanchet:2013haa} and/or numerical studies \cite{Pretorius:2005gq, Campanelli:2005dd, Baker:2005vv, Lehner:2014asa} have been considered; for example, a number of analytic predictions of gravitational waves have been computed within the Post-Newtonian approximation (see e.g. \cite{Blanchet:2013haa} and references therein).

 In this paper we  calculate the gravitational wave signature of two colliding black holes surrounded by a dilatonic field. Such a problem was recently studied numerically for weakly charged black holes where the dilatonic field   vanishes at infinity \cite{Hirschmann:2017psw} and in the Post-Newtonian approximation  for  non-vanishing asymptotic values of the dilaton   \cite{Julie:2017rpw}. (See also \cite{Jai-akson:2017ldo} for a discussion of collisions of dilatonic black holes with angular momentum.)

 We study this problem from a different perspective, analytically calculating the  gravitational wave signature in an approximation that is applicable in the
strong field regime and for any black hole mass ratio. To carry out this procedure it is necessary that the two black holes  be extremally charged and that the system evolve adiabatically, through a series approximately static configurations ---  the so called {\em moduli space approximation} (MSA) \cite{PhysRevLett.59.1617,1989fnr..book.....E_chap}.  We thereby generalize recent results for the Einstein-Maxwell case \cite{Camps:2017gxz},  finding imprints of the dilatonic field on the gravitational wavefront.  As we shall see, such imprints depend crucially on  the value of the dilatonic coupling constant $a$. Interesting analytic results can be obtained at least in two cases: i)   (low energy) string theoretic black holes, characterized by $a=1$, for which there are no coalescence orbits and only a memory effect is observed; and ii)  an intermediate value  $a=1/\sqrt{3}$ of the coupling.  We show that the late-time wavefront in the latter case becomes exponentially suppressed, in notable contrast to  the polynomial decay in the case without a dilaton \cite{Camps:2017gxz}.

 The outline of our paper is as follows. In the next section, we review the evolution of a black hole binary system in the MSA in  Einstein--Maxwell theory. Following \cite{Camps:2017gxz}, the corresponding gravitational wavefront is calculated in Sec. 3. The main results of the paper are gathered in Sec.~4 where the dilatonic case is studied. We conclude in Sec.~5.

\section{Black hole merger in moduli space approximation}

To describe a black hole merger in the MSA in  Einstein--Maxwell theory, we start with   the static multi black hole solution due to {\em Majumdar and Papapetrou} (MP) \cite{PhysRev.72.390, 10.2307/20488481}. The MP solution
represents a static configuration of $n$ extremally charged black holes, each of mass $m_i$ and position $\vx_i$; for $n=1$ it reduces to the  familiar extremal Reissner--Nordstr{\"o}m spacetime. The solution reads
\ba
ds^2&=&-\psi^{-2}dt^2+\psi^2  d\vx \cdot d\vx \,,\label{MP_metric}\\
A&=&-(1-\psi^{-1}) dt\,.\label{MP_gauge_field}
\ea
Here, $A$ is the Maxwell vector potential, and the metric function $\psi$ is given by
\be\label{MP_psi}
\psi=1+\sum_{i=1}^n\frac{m_i}{r_i}\,.
\ee
 In what follows, we shall often write $d\vx\cdot d\vx =dr^2+r^2d\Omega^2$, with $r=\sqrt{\vx \cdot \vx}=|\vx|$. We also have  $r_i = |\vec{r}_i|=\left|\vx-\vx_i\right|$.

The MP solution is static. To describe a dynamical system, we promote
 the black hole positions $\vx_i$ in \eqref{MP_psi} to functions of time, $\vx_i=\vx_i(t)$, and further employ the MSA, requiring that the system moves through configurations with small velocities, always remaining approximately static. This amounts to  perturbing the solution and treating the black holes as slowly moving.
 To second order in velocities one obtains the {\em moduli space metric}, in which the motion of black holes is geodesic
     \cite{PhysRevLett.59.1617,1989fnr..book.....E_chap}. In particular  the following Lagrangian
\be
L=
\frac{1}{2}\mu\gamma(r_{_{12}})\,\vv\cdot\vv \label{moduli_space_Lagrangian}
\ee
describes the centre of mass motion of two black holes, with the centre-of-mass motion  subtracted. Here $M\equiv m_1+m_2$ and $\mu\equiv\frac{m_1m_2}{M}$ are the total and reduced black hole masses, and $\vec r_{_{12}}\equiv \vx_1-\vx_2 $ and $\vv=\frac{d\vec r_{_{12}}}{dt}$ are the relative black hole separation and velocity.
The conformal factor $\gamma(r_{_{12}})$ takes the form
\be
\gamma(r_{_{12}})=\lb1+\frac{M}{r_{_{12}}}\rb^3-\frac{2\mu M^2}{r_{_{12}}^3}.\label{gamma_MP}
\ee
The approximation holds for  \cite{1989fnr..book.....E_chap}
\be\label{large-r}
\frac{r_{_{12}}}{M}\gg v_\infty^2\,,
\ee and so
will certainly break down in the final stages of the black hole coalescence, although note that by choosing small $v_\infty$ we can get arbitrarily close to the complete merger.

All we have to do to study the black hole merger or scattering is to solve the equations of motion
\ba
\dot{\phi}_{_{12}}-\frac{bv_\infty}{r_{_{12}}^2 \gamma(r_{_{12}})}&=&0 \label{EOM_phi}\\
\lb\frac{dr_{_{12}}}{dt}\rb^2+\frac{v_\infty^2}{\gamma(r_{_{12}})}\lb\frac{b^2}{\gamma (r_{_{12}})r_{_{12}}^2}-1\rb&=&0 \label{EOM_MP}
\ea
that  follow from \eqref{moduli_space_Lagrangian}.
 Conservation of energy $E=\frac{1}{2} Mv_\infty^2$ and angular momentum $l=b v_\infty$
follow straightforwardly, with $v_\infty$  the relative velocity at infinite separation of the black holes, and $b$ the impact parameter. Without loss of generality we can confine
the motion  to a plane $\theta=\frac{\pi}{2}$ due to the spherical symmetry of $\gamma(r_{_{12}})$.

These equations of motion allow for both coalescing and scattering orbits, depending on the value of the impact parameter:
if $b>b_{crit}$
, scattering will 
occur, and for $b<b_{crit}$ there will be a merger.
 For any mass ratio, $b_{crit}$ {is obtained by computing the degenerate positive root in the effective potential in \eqref{EOM_MP}, yielding}
\be
 {\frac{2 b_{crit}^3}{3 \sqrt{3}}-b_{crit}^2 M+2 \mu  M^2=0\label{bcrit}}\,,
\ee
which becomes $b_{crit}=\frac{3+\sqrt{3}}{2}M$ for equal mass black holes.

There are two limiting cases of physical interest for which trajectories can be found: i) $M\ll r_{_{12}}$ when the black holes are widely separated (corresponding to early times of the interaction, $t \to -\infty,$ or late times of the black hole scattering, $t\to +\infty$) and ii) $r_{_{12}}\ll M$ for late times for black hole coalescence.

In the first regime, Eqs.  \eqref{EOM_phi} and \eqref{EOM_MP} become
\be
\dot r_{_{12}}=\mp v_{\infty}\bigl(1-\frac{3}{2}\epsilon+\dots\bigr)\,,\quad \dot{\phi}_{_{12}}=\frac{b v_{\infty}\epsilon^2}{M^2}+\dots\,,
\ee
where $\epsilon=M/r_{_{12}}\ll1$, yielding
\ba\label{early_a=0}
r_{_{12}}{}_{early/late}&=& \mp v_{\infty}t-\frac{3}{2}M\log( \mp v_\infty t/r_0)\,,\nonumber\\
\phi_{_{12}}{}_{early/late}&=&-\frac{b}{v_{\infty}t}+\phi_{_{12}}{}_{0}\,,
\ea

For late time coalescing orbits, equations \eqref{EOM_phi} and \eqref{EOM_MP} read
\be
\dot r_{_{12}}=- \frac{v_{\infty}\epsilon^{3/2}\sqrt{M}}{\sqrt{M-2\mu}}\,,\quad
\dot{\phi}_{_{12}}=\frac{b v_{\infty}\epsilon}{M(M-2\mu)}\,,
\ee
where now $\epsilon=r_{_{12}}/M\ll1$, giving
\be\label{late_a=0}
r_{_{12}}{}_{coalescing}= \frac{4M^2(M-2\mu)}{v_\infty^2t^2}\,,\quad
\phi_{_{12}}{}_{coalescing}=-\frac{4b}{v_{\infty}t}\,,
\ee
disregarding the integration constants.

\begin{figure}
\hspace{-2em}
\includegraphics[scale=0.35]{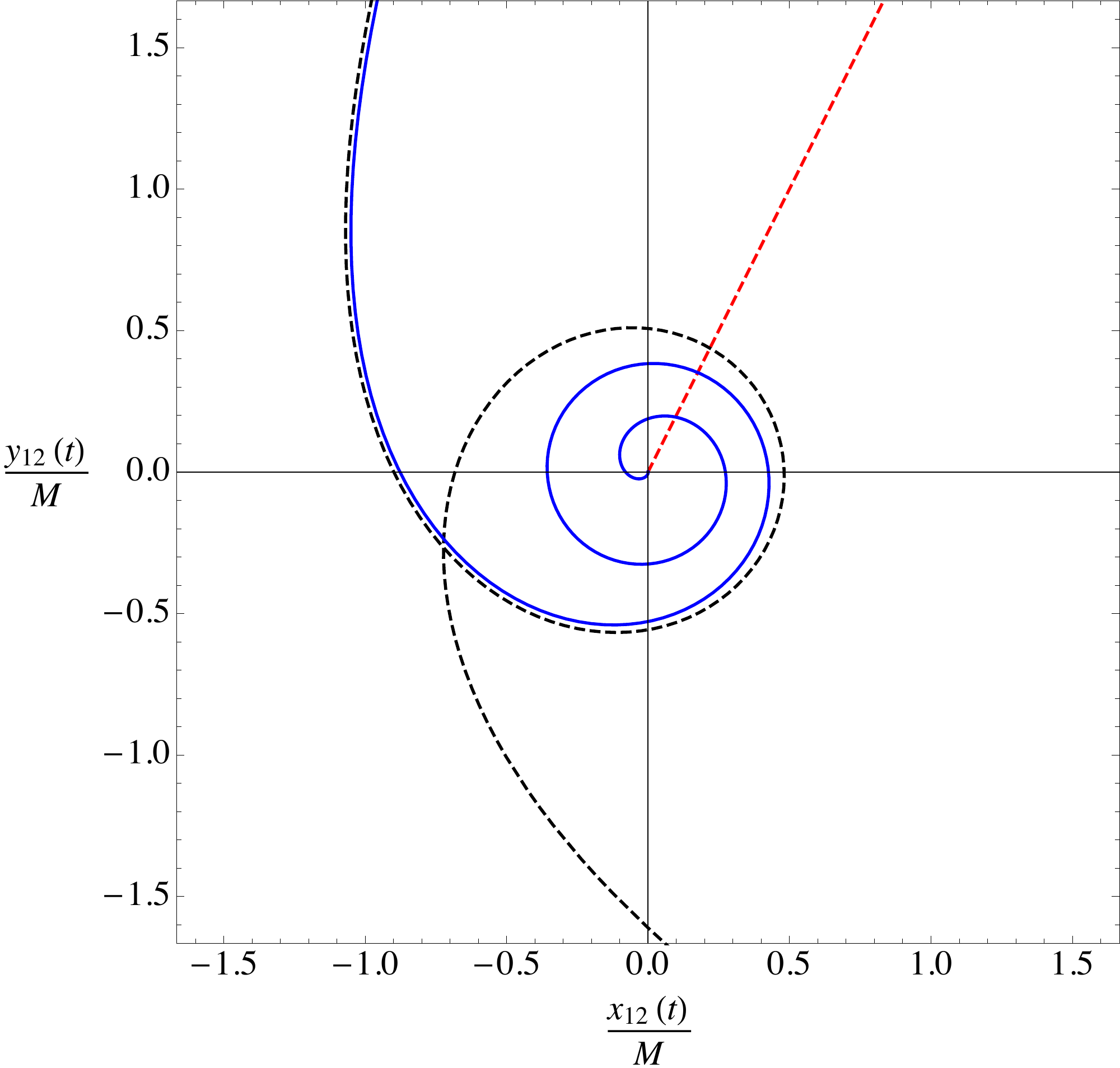}
\caption{ {\bf Black hole trajectories for $a=0$.} Trajectories are illustrated for equal mass black holes and  various impact parameters. In red we have plotted a head-on ($b=0$) collision. The blue solid line corresponds to a slightly-below-critical collision: $b=0.999 b_{crit}=2.36366M$, whereas the black dashed line to a slightly-above-critical scattering: $b=1.01 b_{crit}=2.38969M$. We set up the two near-critical collisions with otherwise identical initial conditions. Recall that $b_{crit}=\frac{3+\sqrt{3}}{2}M\approx2.36603M$. }
\label{Fig:trajectories}
\end{figure}

These simple expressions will allow us to find analytic approximations for the early- and late-time radiation.
For the `intermediate times' we shall solve Eqs. \eqref{EOM_phi} and \eqref{EOM_MP} numerically, to plot the trajectories for various values of $b$. We depict the solutions in Fig.~\ref{Fig:trajectories}, which provides an illustration of trajectories just above and just below the critical impact parameter for a collision.

\section{Gravitational Radiation to Leading-Order }

Following closely the discussion in \cite{Camps:2017gxz}, let us now study the gravitational radiation from the binary black hole system described by the moduli space approximation.

To leading order, gravitational radiation experienced by an observer at radial coordinate $r$ is given by the quadrupole formula
\be
h^{TT}=\frac{2}{r}\frac{d^2}{dt^2}Q^{TT}\bigg{|}_{t_{ret}}\,.
\ee
Here, $h$ is the metric perturbation describing the gravity wave, $Q$ is the mass quadrupole, $TT$ denotes the transverse-traceless projection, and $t_{ret}=t-r$ is the retarded time. For a metric such as \eqref{MP_metric}, it is easy to read off $Q$ due to its \ti{asymptotically Cartesian mass-centred} form (see Section XI of \cite{RevModPhys.52.299} for a definition of this form). In the centre of mass frame, the expansion of $g_{tt}$ gives\footnote{Here, the $Y_{l}\,^m$ are the spherical harmonics normalized such that $\int Y_{l}\,^m\bar{Y}_{l'}\,^{m'}d\Omega=\delta_{l,l'}\delta_{m,m'}$.}
\ba
g_{tt}&=&-1+\frac{2M}{r}+\frac{3 M^2}{r^2}-\frac{4M^3}{r^3}\nn\\
&+&\frac{\mu r_{_{12}}^2}{r^3}\sqrt{\frac{6\pi}{5}}\Bigl(e^{-2 i \phi_{12}}Y_2\,^2-\sqrt{\frac{2}{3}}Y_2\,^0+e^{2i\phi_{12}}Y_2\,^{-2}\Bigr)\nn\\
&+&\mathcal{O}\lb\frac{1}{r^4}\rb\,,\label{expansion_g00}
\ea
where the mass quadrupole moments $I_2\,^m$ are 
\ba
I_2{}^{\pm 2}&=&2\sqrt\frac{2\pi}{5}\mu r{_{_{12}}}^{\!\!\!\!2}e^{\mp 2i\phi_{12}}\,;\quad I_2{}^0=-4\sqrt{\frac{\pi}{15}}\mu r{_{_{12}}}^{\!\!\!\!2} \,,
\ea
obtained by comparing with equation (11.4a) of \cite{RevModPhys.52.299}.
The transverse traceless projection of $Q^{TT}$ is
\ba
Q^{TT}&=&\frac{1}{4}\lb{I}_2\,^2\,_{-2}Y_2\,^2+{I}_2\,^0\,_{-2}Y_2\,^0+{I}_2\,^{-2}\,_{\!-2}Y_2\,^{-2}\rb\hat{e}_R\nonumber\\
&&+c.c.,
\ea
where $c.c.$ stands for complex conjugate, $\hat{e}_R$ is the circular polarisation tensor
\be
\hat{e}_R=\frac{1}{\sqrt{2}}\lb\hat{e}_++i\hat{e}_\times\rb,
\ee
 and $_{-2} Y_l\,^m$ are the spin-weighted spherical harmonics of spin-weight $-2$:
\ba
_{-2}Y_2\,^2&=&\frac{1}{2} \sqrt{\frac{5}{\pi }} e^{2 i \phi } \cos ^4\!\left(\frac{\theta }{2}\right)\,,\quad
_{-2}Y_2\,^0=\frac{1}{4} \sqrt{\frac{15}{2 \pi }} \sin^2\!\theta\,,\nonumber\\
_{-2}Y_2\,^{-2}&=& \frac{1}{2} \sqrt{\frac{5}{\pi }} e^{-2 i \phi } \sin ^4\!\left(\frac{\theta }{2}\right)\,.
\ea
$\lb\theta,\,\phi\rb$ are the angular coordinates of 
the observer. To simplify matters, we can choose an observer on the north pole $\lb\theta,\,\phi\rb=\lb0,0\rb$ (so $_{-2}Y_2\,^0=0=\,_{-2}Y_2\,^{-2}$) and
\be
h^{TT}=\frac{\mu}{\sqrt{2}r}\frac{d^2}{dt^2}\lb r_{_{12}}^2e^{-2i\phi_{12}}\rb\hat{e}_R+c.c.
\ee
 All that remains to calculate the gravitational radiation is to solve \eqref{EOM_MP} for ${r_{_{12}}}$ and ${\phi_{_{12}}}$.
This can easily be done numerically, or, for  early and late times, analytically, using the results of the previous section. Using  \eqref{early_a=0} we find
\be
h_{early/late}^{TT}=\frac{{\sqrt{2}}\mu v_\infty^2}{r}\lb1\pm \frac{3}{2} \frac{M}{v_\infty t}\rb e^{-2 i\phi_{_{12}}}\hat{e}_R+c.c.\,,\label{early_late_htt}
\ee
where the upper/lower signs correspond to {early/late} time scattering orbits.
 As noted in \cite{Camps:2017gxz}, (\ref{early_late_htt}) provides a clear illustration of the {\em gravitational memory effect}: $h^{TT}$ takes different values at early and late times and we have
\be
\Delta h^{TT}=\frac{\sqrt{2}\mu v_\infty^2}{r}\lb e^{-2 i \phi_{12}^{f}}-e^{-2 i \phi_{12}^{i}}\rb
\hat{e}_R+c.c.\tcb{\,,}
\ee
 where $\phi_{_{12}}^i$ and $\phi_{_{12}}^f$ are the respective  initial and final angular separations.
For coalescing orbits at late times \eqref{late_a=0} we recover
\be
h_{coalescing}^{TT}=\frac{{160}\sqrt{2}\mu M^4 (M-2\mu)^2}{r t^6 v_\infty^4}e^{-2i\phi_{12}}\hat{e}_R+c.c.,
\ee
and we note that, at late times of a coalescence, the $t$-dependence of $\phi_{12}$ is too small to appear at this order in $h^{TT}$.
 Note also the  $t^{-6}$ fall-off, characteristic for  Einstein--Maxwell theory.  As we shall see in the next section, this becomes very different in the presence of a dilaton.

\begin{figure}[h!]
\includegraphics[scale=0.8]{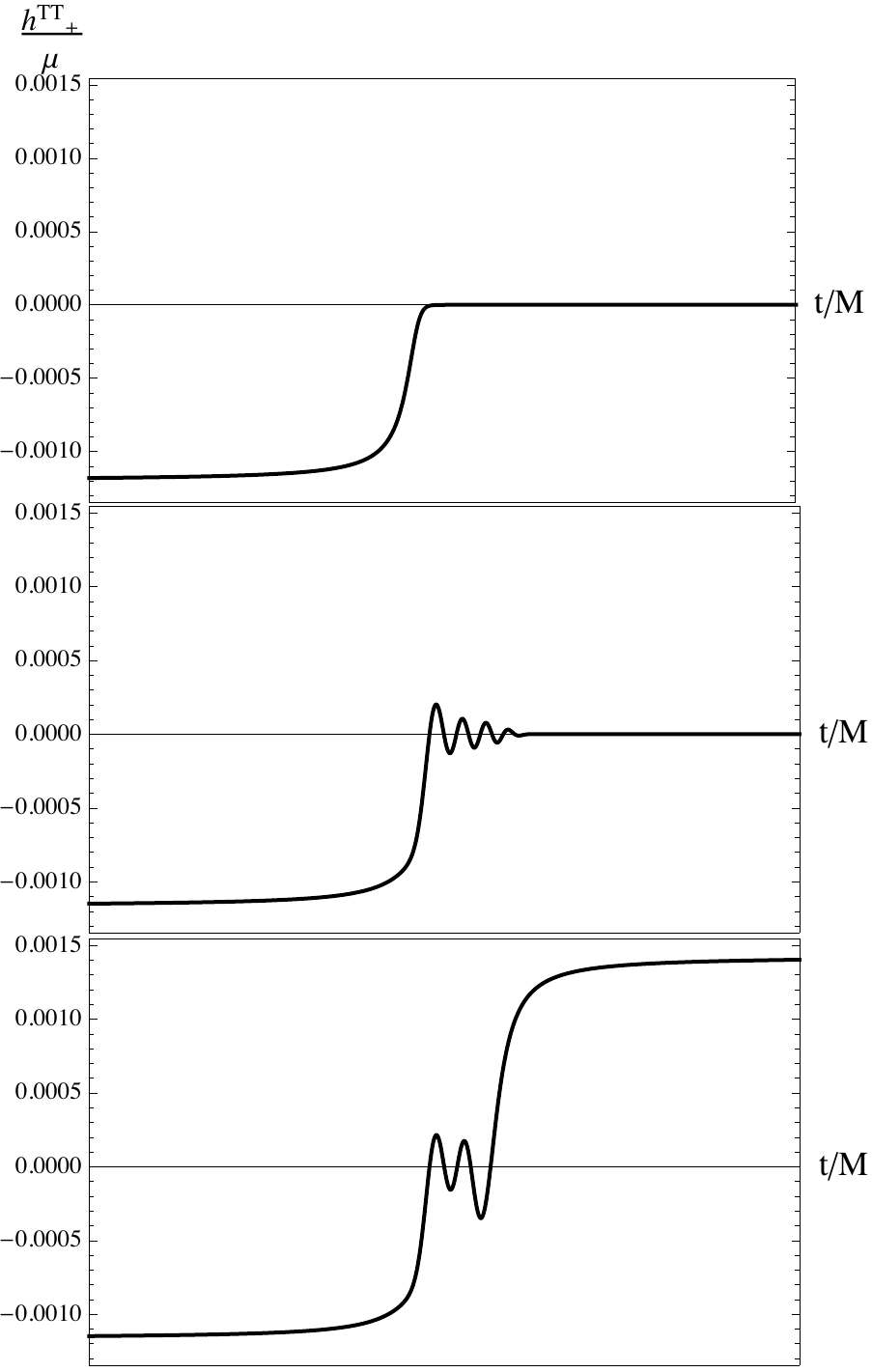}
\caption{ {\bf Gravitational wave signatures for $a=0$.} We have plotted the $h^{TT}_+$ wavefronts for the three different orbits depicted in Fig. \ref{Fig:trajectories}. The top graph illustrates the wavefront emitted upon the head-on ($b=0$) merger, the middle graph the wavefront emitted upon the below-critical coalescence, and the bottom graph the wavefront emitted upon the scattering interaction.
}\label{fig:azero}
\end{figure}

The $h^{TT}_+$ signatures can be seen in Fig.~\ref{fig:azero}, where we have plotted the \ti{numerically calculated} signatures for orbits with impact parameters $b=0$, $b=0.999 b_{crit}$, and $b=1.01 b_{crit}$.  See also Fig. \ref{Fig:log},  where we plot the logarithm of the numerically calculated wavefront for a head-on and a near-critical merger and include the early- and late-time analytic expressions for comparison purposes; the analytic predictions are followed closely.

\section{Coupling to a Dilaton}

 Let us now consider the following generalization of the Einstein--Maxwell theory \cite{GIBBONS1988741,PhysRevD.43.3140}:
\be
S= \int d^4x \sqrt{-g}\lb-R+2\lb\nabla\phi\rb^2+e^{-2a\phi}F^2\rb\,,\label{dilaton_action}
\ee
with $\phi$ a dilatonic scalar field and $a$ the corresponding coupling constant. This action
describes a broad range of fundamental theories:   $a=0$ yields Einstein--Maxwell theory, $a=1$ gives the low energy action of string theory,
 and $a=\sqrt{3}$ corresponds to  Kaluza--Klein theory.
The corresponding static multi-black hole solution \cite{Shiraishi:1992hz} is given by
\ba\label{dilaton_MBH}
ds^2&=&-\psi_a^{-\frac{2}{1+a^2}}dt^2+\psi_a^{\frac{2}{1+a^2}}d\vx\cdot d\vx\,,\nonumber\\
A&=&\frac{1}{\sqrt{1+a^2}}\psi_a^{-1}dt\,,\quad e^{-2a\phi}=\psi_a^{\frac{2a^2}{1+a^2}}\,.\label{dilaton_soln}
\ea
where
\be
\psi_a=1+(1+a^2)\sum_{i=1}^n\frac{m_i}{r_i}\,,\label{dilaton_psi}
\ee
and reduces to the MP solution (\ref{MP_psi}) for $a=0$.

The dilatonic multi-black hole solutions are smooth in the conformal frame\footnote{Such a conformal frame is different from the Jordan frame considered typically in string theory for $a=1$, which would be obtained by $g^{(s)}_{ab}=e^{2a\phi} g_{ab}$. Contrary to the extremal electrically charged multi-black hole solutions \cite{Shiraishi:1992hz} that are regular in the frame $\tilde g_{ab}$, the extremal magnetically charged multi-black hole solutions \cite{PhysRevD.43.3140} are regular in the string frame. }   with metric $\tilde{g}_{ab}=e^{-2a\phi} g_{ab}$ but are singular at the horizon in the Einstein frame with metric  $g_{ab}$ in \eqref{dilaton_MBH}, a point  noted previously  \cite{Shiraishi:1995sd}.  However  in the Einstein frame both the
the moduli space approximation \cite{Shiraishi:1992nq} and an effective field theory \cite{Degura:2000tx} can be fully worked out for general $a$. These approximations are valid provide the black holes are sufficiently separated (eq. \eqref{large-r});  within this context the singular behaviour at the horizons does not affect the motion of these extremal objects.

 In order to find the quadrupole moment for this metric, we need to perform an expansion of $g_{tt}$, similar to  \eqref{expansion_g00},
obtaining in general $a$-dependent coefficients of expansion.  The structures of equation \eqref{dilaton_soln} and \eqref{dilaton_psi} ensure that the quadrupole moment is $a$-independent  and  is still given by \eqref{expansion_g00}.

 Let us now promote the static metric to a dynamical setting, using the MSA approximation.
The corresponding moduli space metric for general $a$ \cite{Shiraishi:1992nq} yields a description of the motion of two black holes via  the Lagrangian (\ref{moduli_space_Lagrangian}) where now{\footnote{{Note that this reduces to equation (IV.9) in ref. \cite{Julie:2017rpw} for the weak field $\psi_a\ra1$ approximation.} }}
\be\label{MS_general_a}
\gamma_a(r_{_{12}})=1+M\bigl(\frac{3-a^2}{4 \pi }\bigr)\int d^3x\bigl(\psi_a^{\frac{2(1-a^2)}{1+a^2}}\bigr)\frac{\vec{r}_1\cdot\vec{r}_2}{r_1^3 r_2^3}\,,
\ee
and is non-trivial to integrate for
generic $a$.  Of course, for $a=0$  we obtain (\ref{gamma_MP}).  For the Kaluza--Klein case, $a=\sqrt{3}$, the moduli space metric vanishes and there is no interaction between the black holes at this order of expansion---to get non-trivial results one would have to go to higher order in velocities.

In what follows we focus on  two cases where we can perform analytically the integration in (\ref{MS_general_a}): the string-theoretic case
 $a=1$  for which \cite{Shiraishi:1992nq}
\be
\gamma_{a=1}=1+\frac{2 M}{r_{_{12}}\,},
\ee
and the case  $a=\frac{1}{\sqrt{3}}$, where we find
\be
\gamma_{a=\frac{1}{\sqrt{3}}}=1+\frac{8}{3}\lb\frac{M}{ r_{_{12}}}+\frac{2M^2}{3r_{_{12}}^2}\rb\,.\label{gamma_root3}
\ee
Note that these are both  independent of $\mu$, in contrast to what happens in the Einstein-Maxwell case.  In other words, in our approximation and for these two special cases the gravitational wave signature will only depend on the total mass of the system but not on the binary mass ratio.

\subsection{String theory black holes: $a=1$}

\begin{figure}[h!]
\includegraphics[scale=0.5]{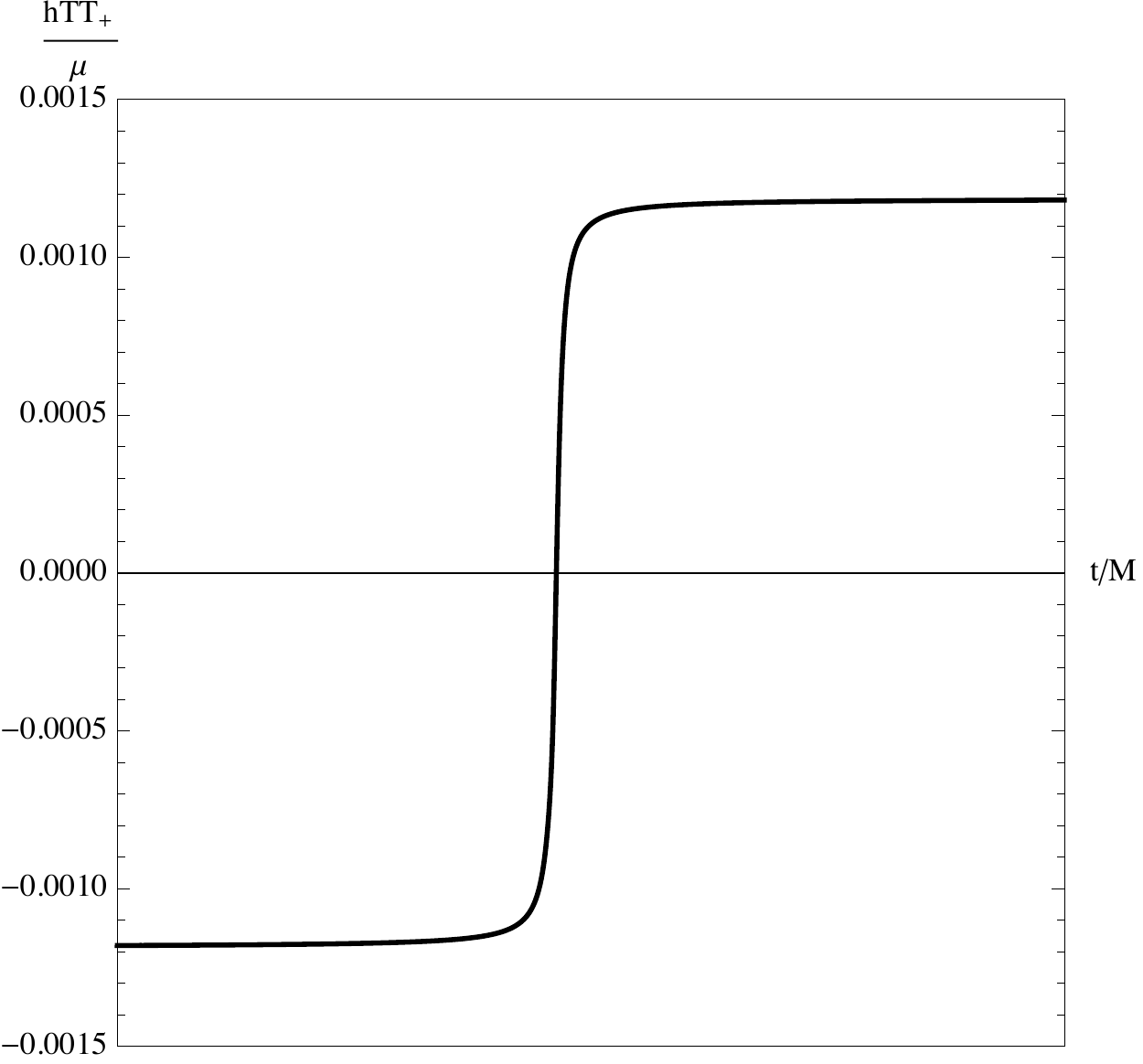}
\caption{ {\bf The memory effect for $a=1$.}  For this case there are no merging orbits and  no  oscillatory behaviour in $h^{TT}$. However we do see very clearly a memory effect.
}\label{fig:a1}
\end{figure}

When $a=1$, $\psi$ does not contribute at all to the integral in (\ref{MS_general_a}).  Interestingly, there is no value of $b$ for which the black holes merge.  At least within the MSA,  all trajectories are scattering, including the head-on collision \cite{Shiraishi:1992nq}  (although it is not unreasonable to suspect that mergers could happen when the approximation is taken to higher order in $v^2$). As such, no oscillatory waveforms exist, and we only observe a memory effect, according to
\ba
 {{\phi}_{_{12}}\,_{early/late}}&=& -\frac{b}{v_\infty t}+\dots\,,\\
r_{_{12}}\,_{early/late}&=&\mp v_\infty t- M \log \lb\mp v_\infty t\rb+\dots\,,
\ea
and so
\be\label{htta1}
h_{early/late}^{TT}=\frac{\sqrt{2}\mu v_\infty^2}{r}\lb1\pm \frac{ M }{ v_\infty t}\rb e^{-2 i\phi_{12}}\hat{e}_R+c.c.\,.
\ee
The memory effect can be seen in Figure \ref{fig:a1}. 

\begin{figure}[h!]
\includegraphics[scale=0.5]{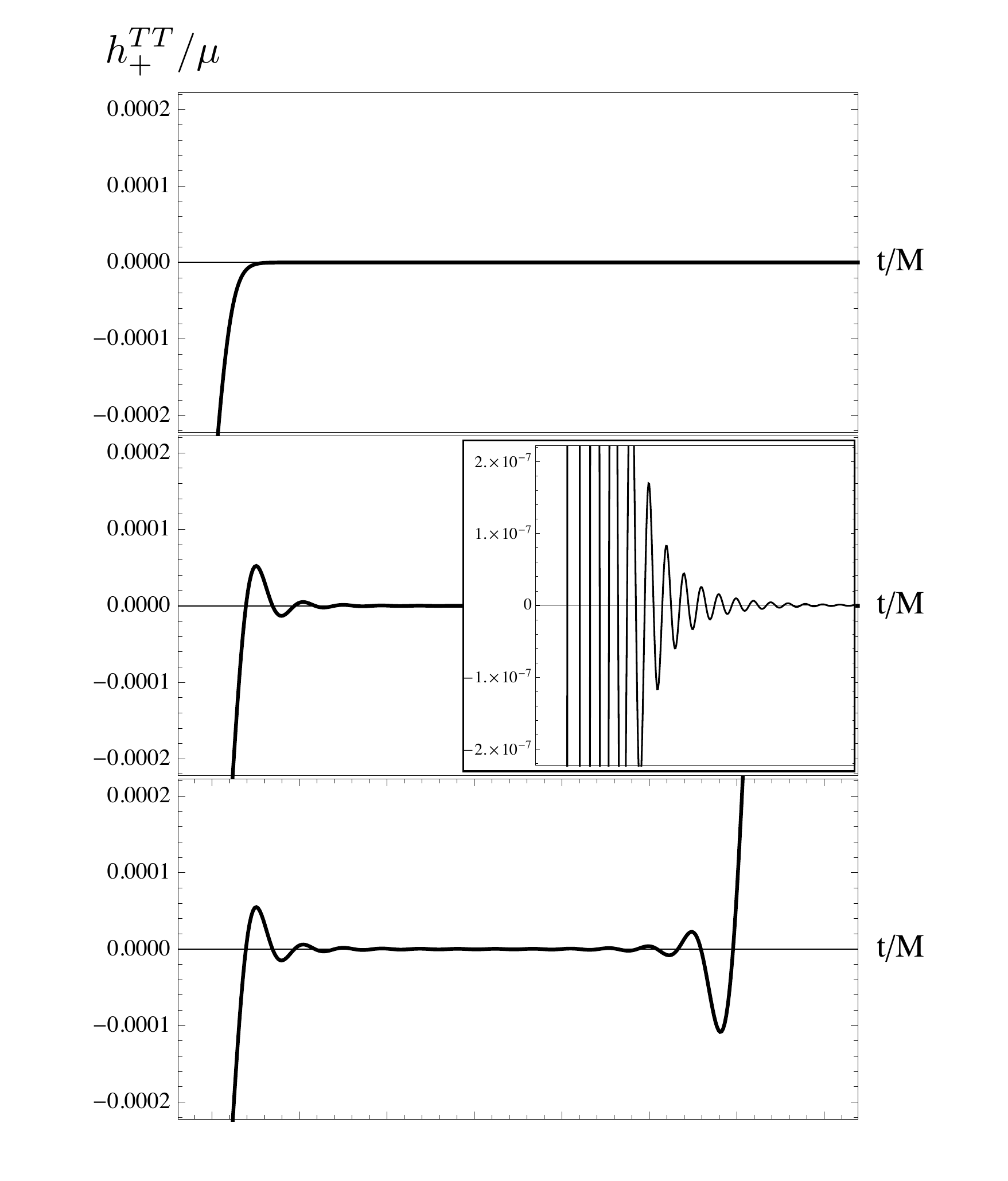}
\caption{ {\bf Gravitational wave signatures for $a=\frac{1}{\sqrt{3}}$.} We plot here graphs
analagous  to those in Figure \ref{fig:azero}. The top graph illustrates the wavefront emitted upon a head-on collision ($b=0$),  the middle  a sub-critical case ($b=0.999 b_{crit}=1.332M$), and the bottom   a scattering event ($b=1.01 b_{crit}=1.34667M$). The inset in the middle depicts  near-critical coalescence to make the exponentially decaying behaviour more explicit.}

\label{fig:athree}

\end{figure}

\subsection{Intermediate coupling: $a=\frac{1}{\sqrt{3}}$}

For $a=\frac{1}{\sqrt{3}}$ the $\mu$-independence of  $\gamma(r_{_{12}})$  in (\ref{gamma_root3})
 implies that wavefronts emitted by binary pairs of arbitrary mass ratio yield the same gravitational wave signature, albeit rescaled by $\mu$; this is not true for $a=0$, for which the equations of motion explicitly depend on $\mu$. The critical impact parameter $b_{crit}$ is $b_{crit}=\frac{4}{3}M$.

\bfig[h!]

\includegraphics[scale=0.45]{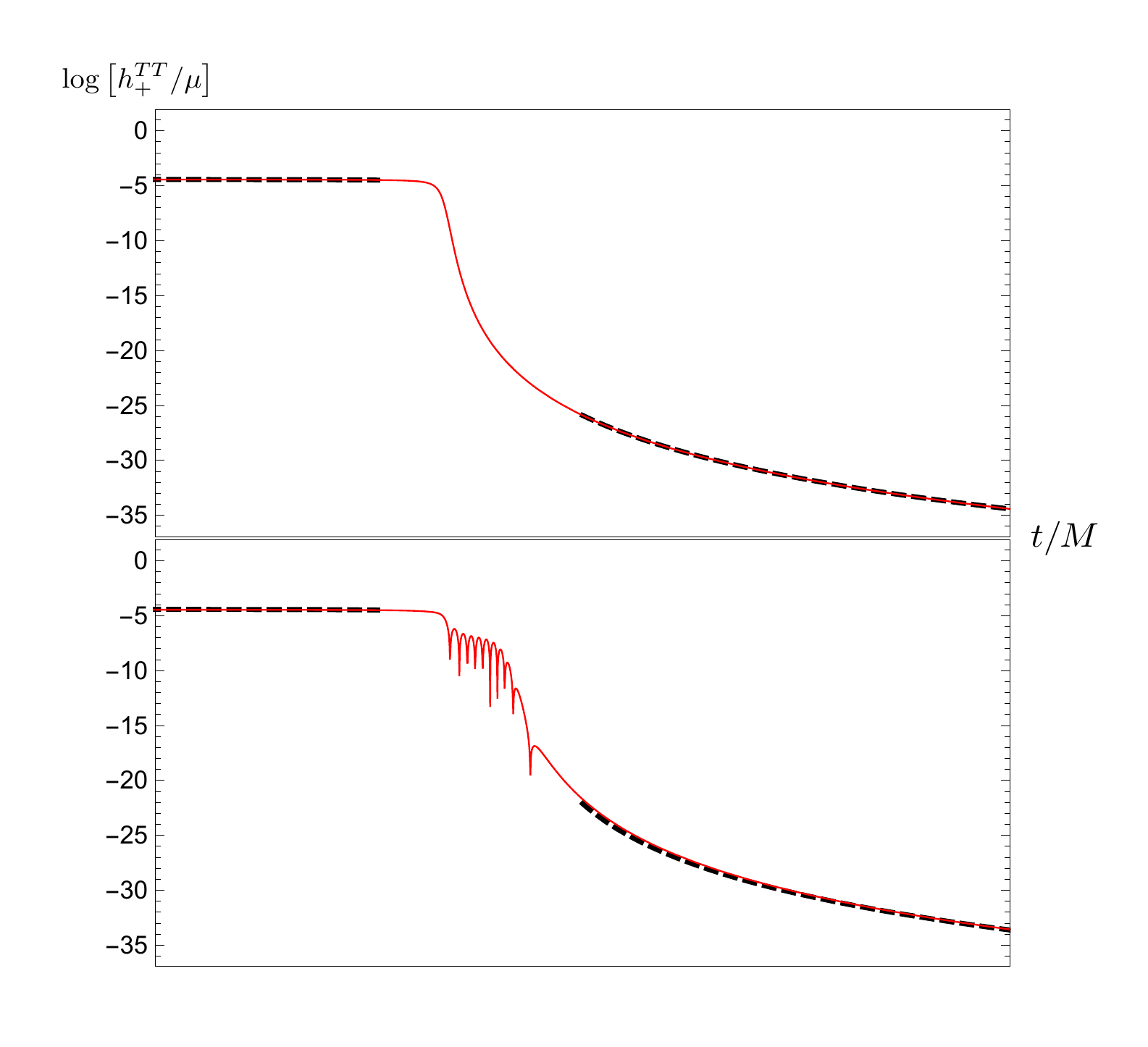}

\centering{(a) $a=0$, with (above) $b=0$ and (below) $b=0.999 b_{crit}$.}
\includegraphics[scale=0.45]{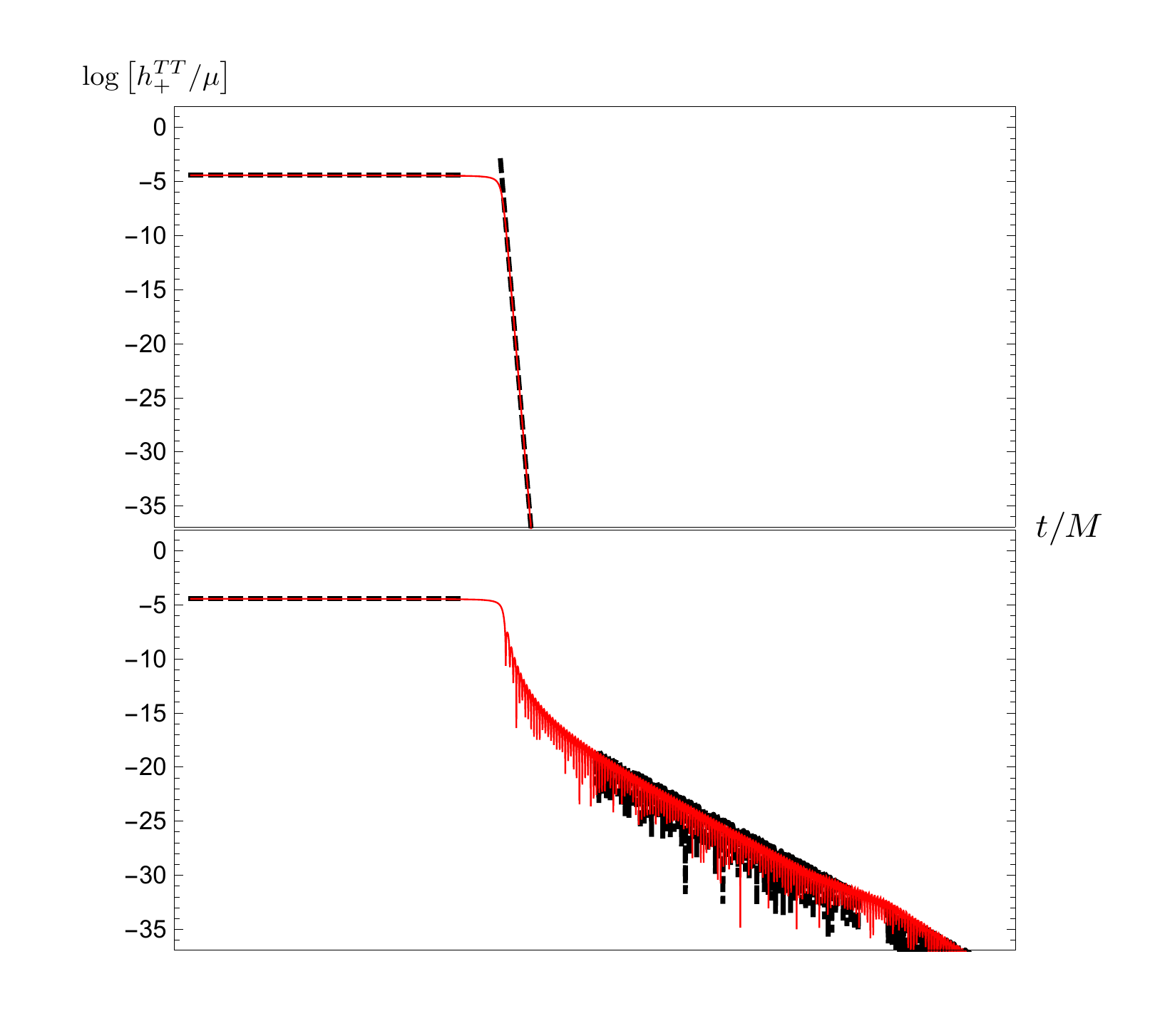}

\centering{(b) $a=\frac{1}{\sqrt{3}}$, with (above) $b=0$ and (below) $b=0.999 b_{crit}$.}
\caption{{\bf Comparison of analytic expressions with numerical results for $a=0$.} Above we have plotted the behaviour $\log(\left|h^{TT}_+\right|)$ as a function of time for head-on and below-critical collisions, with the numerically calculated solution in red and the analytical predictions for large and small $\frac{m}{r_{_{12}}}$ in black. (a) is for $a=0$ and (b) is for $a=\frac{1}{\sqrt{3}}$. We notice a number of things on the log plot that are difficult to see on the previous graphs: the $t^{-6}$ for $a=0$ behaviour can be directly contrasted with the $e^{-t}$ behaviour for $a=\frac{1}{\sqrt{3}}$. We also notice the lack of $b$-dependence for the $a=0$ case, as predicted, and the obvious $b$-dependence for $a=\frac{1}{\sqrt{3}}$.}
\label{Fig:log}
\efig

Using (\ref{gamma_root3}) to solve (\ref{EOM_MP})  yields
 \ba
r_{_{12}}\,_{early/late}&=&\mp v_\infty t-\frac{4M}{3}\log\lb\mp v_\infty t\rb+\dots\,,\\
 {\phi_{12}\,_{early/late}}&=& -\frac{b}{v_\infty t}+\dots\,,\label{early_phi_root3}
\ea
for the separation at early and late times, when $r_{_{12}}\gg M$.
Hence
\be
h^{TT}_{early/late}=\frac{ \sqrt{2}\mu v_\infty^2}{r}\lb1\pm \frac{4}{3}\frac{M}{v_\infty t}\rb e^{-2i\phi_{_{12}}}\hat{e}_R+c.c.\,.
\ee

For coalescing orbits at late times we find
\ba
r_{_{12}}\,_{coalescing}&=&r_{0} \exp\bigl(-\frac{3q}{16}\frac{v_\infty t}{M}\bigr)\,,\label{rcoalescing3}\\
 {{\phi}_{_{12}}\,_{coalescing}}&=& \frac{9}{16} \frac{b v_\infty t}{M^2}+\dots\,\label{phicoalescing3},
\ea
where $r_{0}$ is the separation at some $t_0$, and we abbreviated $q\equiv\sqrt{16-9b^2/M^2}$
; in particular, note that $\dot{\phi}_{_{12}}$ is no longer small at late times. This implies an exponentially decaying signature:
\ba
{h^{TT}_{coalescence}}&=&\frac{9}{64}\frac{\mu\sqrt{2}r_{0}^2v_\infty^2}{rM^4}(8M^2-9b^2+3ibMq)\nonumber\\
&\times& \exp\Big{(}-\frac{3}{8}\frac{v_\infty t}{M^2}(Mq+3ib)\Big{)}\hat{e}_R+c.c.\,.\quad
\ea

 We show the logarithm of the wavefront of coalescing orbits for different values of $b$ in Figure \ref{Fig:log}(b), where the $b$-dependence of the fall-off is seen. The exponential fall-off behaviour is also clearly shown, in contrast to the $t^{-6}$ behaviour evident in Figure \ref{Fig:log} (a) for the Einstein--Maxwell case.
{Note that the electromagnetic radiation would also be expected to have an exponential fall-off, as it takes a similar form as gravitational radiation (see \cite{Camps:2017gxz}).}

\section{Conclusion}

The presence of a dilaton  can make a significant imprint on the gravitational waveforms
emitted in black hole collisions and scattering events.
 By analytically computing expressions for the gravitational wavefronts emitted by the collision of two extremally charged dilatonic black holes, we have been able to compare
the general relativistic (Einstein--Maxwell)  wavefronts  with those occurring in a string-theoretic case ($a=1$) and a more general  dilatonic theory ($a=1/\sqrt{3}$ ).  In the latter case the gravitational waveforms are exponentially suppressed in time, whereas in general relativity the wavefronts decay with $t^{-6}$.  However the gravitational memory effect for   scattering is the same for all values of $a$, including the $a=0$ Einstein--Maxwell case.

  Our results complement those of recent  studies of dilatonic black hole mergers \cite{Hirschmann:2017psw, Julie:2017rpw}, and  illustrate a qualitative difference between cases with and without a dilaton.

It would be interesting to develop this technique to spacetimes with general coupling constant $a$ between the dilaton and the Maxwell field as we have only been able to do this so far for the  specific  values of $a=0,\frac{1}{\sqrt{3}}, 1, \sqrt{3}$; we leave this question for a future study. Likewise, more detailed studies of non-extremal dilatonic black holes over a broad range of parameter space need to be carried out to see where the most interesting phenomenological possibilities lie.

\section*{Acknowledgements}
This work was supported in part by the Natural
Sciences and Engineering Research Council of Canada and by the Perimeter Institute for Theoretical
Physics. Research
at Perimeter Institute is supported by the Government
of Canada through the Department of Innovation,
Science and Economic Development Canada and by the
Province of Ontario through the Ministry of Research,
Innovation and Science.

%

\end{document}